
%
%
%

\message
{JNL.TEX version 0.92 as of 6/9/87.}

\catcode`@=11
\expandafter\ifx\csname inp@t\endcsname\relax\let\inp@t=\input
\def\input#1 {\expandafter\ifx\csname #1IsLoaded\endcsname\relax
\inp@t#1%
\expandafter\def\csname #1IsLoaded\endcsname{(#1 was previously loaded)}
\else\message{\csname #1IsLoaded\endcsname}\fi}\fi
\catcode`@=12



\font\twelverm=cmr10 scaled 1200    \font\twelvei=cmmi10 scaled 1200
\font\twelvesy=cmsy10 scaled 1200   \font\twelveex=cmex10 scaled 1200
\font\twelvebf=cmbx10 scaled 1200   \font\twelvesl=cmsl10 scaled 1200
\font\twelvett=cmtt10 scaled 1200   \font\twelveit=cmti10 scaled 1200
\font\twelvesc=cmcsc10 scaled 1200  
\skewchar\twelvei='177   \skewchar\twelvesy='60


\def\twelvepoint{\normalbaselineskip=12.4pt plus 0.1pt minus 0.1pt
  \abovedisplayskip 12.4pt plus 3pt minus 9pt
  \belowdisplayskip 12.4pt plus 3pt minus 9pt
  \abovedisplayshortskip 0pt plus 3pt
  \belowdisplayshortskip 7.2pt plus 3pt minus 4pt
  \smallskipamount=3.6pt plus1.2pt minus1.2pt
  \medskipamount=7.2pt plus2.4pt minus2.4pt
  \bigskipamount=14.4pt plus4.8pt minus4.8pt
  \def\rm{\fam0\twelverm}          \def\it{\fam\itfam\twelveit}%
  \def\sl{\fam\slfam\twelvesl}     \def\bf{\fam\bffam\twelvebf}%
  \def\mit{\fam 1}                 \def\cal{\fam 2}%
  \def\sc{\twelvesc}               \def\tt{\twelvett}
  \def\sf{\twelvesf}
  \textfont0=\twelverm   \scriptfont0=\tenrm   \scriptscriptfont0=\sevenrm
  \textfont1=\twelvei    \scriptfont1=\teni    \scriptscriptfont1=\seveni
  \textfont2=\twelvesy   \scriptfont2=\tensy   \scriptscriptfont2=\sevensy
  \textfont3=\twelveex   \scriptfont3=\twelveex  \scriptscriptfont3=\twelveex
  \textfont\itfam=\twelveit
  \textfont\slfam=\twelvesl
  \textfont\bffam=\twelvebf \scriptfont\bffam=\tenbf
  \scriptscriptfont\bffam=\sevenbf
  \normalbaselines\rm}



\def\beginlinemode{\endmode
  \begingroup\parskip=0pt \obeylines\def\\{\par}\def\endmode{\par\endgroup}}
\def\beginparmode{\endmode\rightskip=0 pt \leftskip=0pt
  \begingroup \def\endmode{\par\endgroup}}
\let\endmode=\par
{\obeylines\gdef\
{}}
\def\singlespace{\baselineskip=\normalbaselineskip}

\def\oneandahalfspace{\baselineskip=\normalbaselineskip
  \multiply\baselineskip by 3 \divide\baselineskip by 2}
\def\doublespace{\baselineskip=\normalbaselineskip \multiply\baselineskip by 2}

\newcount\firstpageno
\firstpageno=2
\footline={\ifnum\pageno<\firstpageno{\hfil}\else{\hfil\twelverm\folio\hfil}\fi}

\def\toppageno{\global\footline={\hfil}\global\headline
  ={\ifnum\pageno<\firstpageno{\hfil}\else{\hfil\twelverm\folio\hfil}\fi}}
\let\rawfootnote=\footnote              
\def\footnote#1#2{{\rm\doublespace\parindent=0pt\parskip=0pt
  \rawfootnote{#1}{#2\hfill
  \vrule height 0pt depth 6pt width 0pt}}}
\def\raggedcenter{\leftskip=4em plus 12em \rightskip=\leftskip
  \parindent=0pt \parfillskip=0pt \spaceskip=.3333em \xspaceskip=.5em
  \pretolerance=9999 \tolerance=9999
  \hyphenpenalty=9999 \exhyphenpenalty=9999 }
\def\dateline{\rightline{\ifcase\month\or
  January\or February\or March\or April\or May\or June\or
  July\or August\or September\or October\or November\or December\fi
  \space\number\year}}
\def\received{\vskip 3pt plus 0.2fill
 \centerline{\rm Received \ \ifcase\month\or
  January\or February\or March\or April\or May\or June\or
  July\or August\or September\or October\or November\or December\fi
  \qquad,\ \number\year}}


\hsize=6.5truein
\hoffset=0truein
\vsize=8.9truein
\voffset=0truein
\parskip=0 pt
\rightskip=0 pt
\leftskip=0pt
\parindent 1cm
\def\\{\cr}
\twelvepoint            
\doublespace            
\overfullrule=0pt       


      \newcount\timehour
      \newcount\timeminute
      \newcount\timehourminute
      \def\daytime{\timehour=\time\divide\timehour by 60
        \timehourminute=\timehour\multiply\timehourminute by-60
        \timeminute=\time\advance\timeminute by \timehourminute
        \number\timehour:\ifnum\timeminute<10{0}\fi\number\timeminute}
      \def\today{\number\day\space\ifcase\month\or Jan\or Feb\or Mar
           \or Apr\or May\or Jun\or Jul\or Aug\or Sep\or Oct\or
           Nov\or Dec\fi\space\number\year}


\def\title                      
  {\null\vskip 3pt plus 0.2fill
   \beginlinemode \doublespace \raggedcenter \bf}

\def\author                     
  {\vskip 3pt plus 0.2fill \beginlinemode
   \singlespace \raggedcenter\sc}

\def\affil                      
  {\vskip 3pt plus 0.1fill \beginlinemode
   \oneandahalfspace \raggedcenter \it}

\def\abstract                   
  {\vskip 3pt plus 0.3fill \beginparmode \narrower
\oneandahalfspace ABSTRACT:}

\def\endtitlepage               
  {\smallskip\endpage                     
   \body}

\def\body                       
  {\beginparmode}               

\def\head#1{                    
  \goodbreak\vskip 0.2truein    
  {\immediate\write16{#1}
   \raggedcenter \uppercase{#1}\par}
   \nobreak\vskip 0.1truein\nobreak}

\def\beginitems{
\par\medskip\bgroup\def\i##1 {\item{##1}}\def\ii##1 {\itemitem{##1}}
\leftskip=36pt\parskip=0pt}
\def\enditems{\par\egroup}

\def\beneathrel#1\under#2{\mathrel{\mathop{#2}\limits_{#1}}}


\def\references                 
  {\vfill\break
   \head{REFERENCES}            

   \beginparmode
   \frenchspacing \parindent=0pt \leftskip=1truecm
   \parskip=0pt plus 3pt \everypar{\hangindent=\parindent}}

\gdef\refis#1{\indent\hbox to 0pt{\hss#1.~~}
                        \ignorespaces}            

\gdef\journal#1, #2, #3, 1#4#5#6{\ {\sl #1~}{\bf #2}, #3 (1#4#5#6)}

\def\refstyleapj{                
  \gdef\journal##1, ##2, ##3.{             
     {##1},{\ ##2},\ ##3.}
  \gdef\abstract {              
     \vskip 3pt plus 0.3fill
     {\raggedcenter ABSTRACT}
     \beginparmode
     \doublespace}
  \gdef\subjectheadings {\vskip 3 pt plus 0.1fill \beginparmode \noindent
                        {\it Subject headings}:\ }
  \gdef\received{\vskip .75cm
   \centerline{\it Received \number\year\ \ifcase\month\or
    January\or February\or March\or April\or May\or June\or
    July\or August\or September\or October\or November\or December\fi
    \qquad;\ accepted\ \number\year\qquad\qquad\qquad}}
  \gdef\affil                      
    {\beginlinemode
    \oneandahalfspace \raggedcenter}
  \gdef\author                     
    {\vskip 15pt \beginlinemode
    \singlespace \raggedcenter\sc}
  \gdef\references {\vfill\break\head{REFERENCES} \beginparmode
               \parskip=0pt plus .1fil
               \parindent=0pt\frenchspacing\doublespace
               \everypar{\hangindent=.5cm\hangafter=1}}
  \gdef\postaladdresses{\vfill\eject\head{POSTAL ADDRESSES}\beginparmode
                        \parindent=0pt\parskip=0pt plus .1fil}
  \gdef\address##1: ##2.{\hangindent=.5cm\hangafter=1{\sc ##1}: ##2}
}

\def\figurecaptions             
  {\endpage
   \beginparmode
   \head{Figure Captions}
}

\def\endpage                    
  {\vfill\eject}

\def\endpaper                   
  {\endmode\vfill\supereject}


\def\heading                            
  {\vskip 0.5truein plus 0.1truein      
   \beginparmode \def\\{\par} \parskip=0pt \singlespace \raggedcenter}

\def\subheading                         
  {\vskip 0.25truein plus 0.1truein     
   \beginlinemode \singlespace \parskip=0pt \def\\{\par}\raggedcenter}

\def\tag#1$${\eqno(#1)$$}

\def\align#1$${\eqalign{#1}$$}

\def\aligntag#1$${\gdef\tag##1\\{&(##1)\cr}\eqalignno{#1\\}$$
  \gdef\tag##1$${\eqno(##1)$$}}

\def\endaligntag{}

\def\overset #1\to#2{{\mathop{#2}\limits^{#1}}}
\def\underset#1\to#2{{\let\next=#1\mathpalette\undersetpalette#2}}
\def\undersetpalette#1#2{\vtop{\baselineskip0pt
\ialign{$\mathsurround=0pt #1\hfil##\hfil$\crcr#2\crcr\next\crcr}}}


\def\ref#1{Ref.~#1}                     
\def\[#1]{[\call{#1}]}
\def\cite#1{{#1}}
\def\(#1){(\call{#1})}
\def\call#1{{#1}}
\def\taghead#1{}
\def\frac#1#2{{#1 \over #2}}

\def\12{{1\over2}}

\def\sla{\raise.15ex\hbox{$/$}\kern-.57em}
\def\leaderfill{\leaders\hbox to 1em{\hss.\hss}\hfill}
\def\twiddle{\lower.9ex\rlap{$\kern-.1em\scriptstyle\sim$}}
\def\bigtwiddle{\lower1.ex\rlap{$\sim$}}
\def\gtwid{\mathrel{\raise.3ex\hbox{$>$\kern-.75em\lower1ex\hbox{$\sim$}}}}
\def\ltwid{\mathrel{\raise.3ex\hbox{$<$\kern-.75em\lower1ex\hbox{$\sim$}}}}
\def\square{\kern1pt\vbox{\hrule height 1.2pt\hbox{\vrule width 1.2pt\hskip 3pt
   \vbox{\vskip 6pt}\hskip 3pt\vrule width 0.6pt}\hrule height 0.6pt}\kern1pt}
\def\tdot#1{\mathord{\mathop{#1}\limits^{\kern2pt\ldots}}}

\def\pmb#1{\setbox0=\hbox{#1}%
  \kern-.025em\copy0\kern-\wd0
  \kern  .05em\copy0\kern-\wd0
  \kern-.025em\raise.0433em\box0 }

\catcode`@=11
\newcount\tagnumber\tagnumber=0

\immediate\newwrite\eqnfile
\newif\if@qnfile\@qnfilefalse
\def\write@qn#1{}
\def\writenew@qn#1{}
\def\w@rnwrite#1{\write@qn{#1}\message{#1}}
\def\@rrwrite#1{\write@qn{#1}\errmessage{#1}}

\def\taghead#1{\gdef\t@ghead{#1}\global\tagnumber=0}
\def\t@ghead{}

\expandafter\def\csname @qnnum-3\endcsname
  {{\t@ghead\advance\tagnumber by -3\relax\number\tagnumber}}
\expandafter\def\csname @qnnum-2\endcsname
  {{\t@ghead\advance\tagnumber by -2\relax\number\tagnumber}}
\expandafter\def\csname @qnnum-1\endcsname
  {{\t@ghead\advance\tagnumber by -1\relax\number\tagnumber}}
\expandafter\def\csname @qnnum0\endcsname
  {\t@ghead\number\tagnumber}
\expandafter\def\csname @qnnum+1\endcsname
  {{\t@ghead\advance\tagnumber by 1\relax\number\tagnumber}}
\expandafter\def\csname @qnnum+2\endcsname
  {{\t@ghead\advance\tagnumber by 2\relax\number\tagnumber}}
\expandafter\def\csname @qnnum+3\endcsname
  {{\t@ghead\advance\tagnumber by 3\relax\number\tagnumber}}

\def\equationfile{%
  \@qnfiletrue\immediate\openout\eqnfile=\jobname.eqn%
  \def\write@qn##1{\if@qnfile\immediate\write\eqnfile{##1}\fi}
  \def\writenew@qn##1{\if@qnfile\immediate\write\eqnfile
    {\noexpand\tag{##1} = (\t@ghead\number\tagnumber)}\fi}
}

\def\callall#1{\xdef#1##1{#1{\noexpand\call{##1}}}}
\def\call#1{\each@rg\callr@nge{#1}}

\def\each@rg#1#2{{\let\thecsname=#1\expandafter\first@rg#2,\end,}}
\def\first@rg#1,{\thecsname{#1}\apply@rg}
\def\apply@rg#1,{\ifx\end#1\let\next=\relax%
\else,\thecsname{#1}\let\next=\apply@rg\fi\next}

\def\callr@nge#1{\calldor@nge#1-\end-}
\def\callr@ngeat#1\end-{#1}
\def\calldor@nge#1-#2-{\ifx\end#2\@qneatspace#1 %
  \else\calll@@p{#1}{#2}\callr@ngeat\fi}
\def\calll@@p#1#2{\ifnum#1>#2{\@rrwrite{Equation range #1-#2\space is bad.}
\errhelp{If you call a series of equations by the notation M-N, then M and
N must be integers, and N must be greater than or equal to M.}}\else%
 {\count0=#1\count1=#2\advance\count1
by1\relax\expandafter\@qncall\the\count0,%

  \loop\advance\count0 by1\relax%
    \ifnum\count0<\count1,\expandafter\@qncall\the\count0,%
  \repeat}\fi}

\def\@qneatspace#1#2 {\@qncall#1#2,}
\def\@qncall#1,{\ifunc@lled{#1}{\def\next{#1}\ifx\next\empty\else
  \w@rnwrite{Equation number \noexpand\(>>#1<<) has not been defined yet.}
  >>#1<<\fi}\else\csname @qnnum#1\endcsname\fi}

\let\eqnono=\eqno
\def\eqno(#1){\tag#1}
\def\tag#1$${\eqnono(\displayt@g#1 )$$}

\def\aligntag#1\endaligntag
  $${\gdef\tag##1\\{&(##1 )\cr}\eqalignno{#1\\}$$
  \gdef\tag##1$${\eqnono(\displayt@g##1 )$$}}

\def\eqalignno#1{\displ@y \tabskip\centering
  \halign to\displaywidth{\hfil$\displaystyle{##}$\tabskip\z@skip
    &$\displaystyle{{}##}$\hfil\tabskip\centering
    &\llap{$\displayt@gpar##$}\tabskip\z@skip\crcr
    #1\crcr}}

\def\displayt@gpar(#1){(\displayt@g#1 )}

\def\displayt@g#1 {\rm\ifunc@lled{#1}\global\advance\tagnumber by1
        {\def\next{#1}\ifx\next\empty\else\expandafter
        \xdef\csname @qnnum#1\endcsname{\t@ghead\number\tagnumber}\fi}%
  \writenew@qn{#1}\t@ghead\number\tagnumber\else
        {\edef\next{\t@ghead\number\tagnumber}%
        \expandafter\ifx\csname @qnnum#1\endcsname\next\else
        \w@rnwrite{Equation \noexpand\tag{#1} is a duplicate number.}\fi}%
  \csname @qnnum#1\endcsname\fi}

\def\ifunc@lled#1{\expandafter\ifx\csname @qnnum#1\endcsname\relax}

\let\@qnend=\end\gdef\end{\if@qnfile
\immediate\write16{Equation numbers written on []\jobname.EQN.}\fi\@qnend}

\catcode`@=12


\font\fourteenbf=cmbx10 scaled\magstep 2
\def\doublespace{\singlespace}
\doublespace
\twelverm
\vskip 1cm

\centerline{\fourteenbf The Relaxation Effect in Dissipative
Relativistic Fluid Theories\footnote{*}{To Appear in {\sl Annals of Physics}
1996.}}

\vskip 1cm
\centerline{\twelvebf Lee Lindblom}
\vskip 12pt
\centerline{\twelveit Department of Physics, Montana State University}
\centerline{\twelveit {\twelverm and,} D.A.R.C., Observatoire de Paris--Meudon}
\vskip 1cm

\midinsert
\narrower\noindent Abstract:  The dynamics of the fluid
fields in a large class of causal dissipative fluid
theories is studied.  It is shown that the physical fluid states in
these theories must relax (on a time scale that is characteristic of
the microscopic particle interactions) to ones that are essentially
indistinguishable from the simple relativistic Navier-Stokes
descriptions of these states.  Thus, for example, in the relaxed form
of a physical fluid state the stress energy tensor is in effect
indistinguishable from a perfect fluid stress tensor plus small
dissipative corrections proportional to the shear of the fluid
velocity, the gradient of the temperature, etc.
\endinsert

\vskip 1cm
\noindent \S I Introduction
\vskip .5cm

A simple mathematical model provides an elegant and accurate
description of the common materials called fluids.  The effects of
internal dissipation in these materials---viscosity and thermal
conductivity---are also well modeled by a simple generalization of the
basic theory called the Navier-Stokes equations.  Unfortunately, the
most straightforward approaches to constructing relativistic
generalizations of the Navier-Stokes equations result in rather
pathological theories (Eckart [1], Landau and Lifschitz [2]).  These
theories are non-causal, unstable, and without a well posed initial
value formulation (see for example Hiscock and Lindblom [3]).  Less
straightforward approaches have succeeded more recently in producing a
class of causal dissipative fluid theories (e.g., Israel and Stewart
[4], Carter~[5], Liu, M\"uller, and Ruggeri [6], Geroch and Lindblom
[7], etc.).  These theories have eliminated the pathologies of the
straightforward relativistic generalizations of the Navier-Stokes
equations, but they do so at the expense of increasing significantly
the number of dynamical fields needed to describe the fluid.
Unfortunately the additional dynamical degrees of freedom associated
with these extra fields have never been directly observed in real
fluids.  This is probably why these new theories have not found
widespread acceptance.

In this paper the dynamics associated with these additional fluid
fields are analyzed in a very large class of causal dissipative fluid
theories.  It is shown that the physical fluid states relax (on a time
scale characteristic of the inter-particle interactions) to ones that
are also well described by the simple relativistic Navier-Stokes
theory.  For example, the stress-energy tensor in such a relaxed fluid
state is well described by the usual perfect fluid stress-energy
tensor plus the Navier-Stokes expressions for the dissipative
corrections involving the shear of the fluid velocity, the gradient of
the temperature, etc.  This result suggests that meaningful
differences between the causal theories and the non-causal
Navier-Stokes theory can not be observed.  The complicated dynamical
structure of the causal theories is necessary to insure that the fluid
evolves in a causal and stable way.  But this rich dynamical structure
is unobservable, since the physical states of a fluid always evolve in
a way that is also well described by the Navier-Stokes expressions for
the stress-energy tensor, etc.  The arguments which lead to these
conclusions are extremely general: they are based on a fully
non-linear analysis of the equations and do not assume that the fluid
state is close to equilibrium.  This  analysis generalizes
significantly the studies of the analogous relaxation effect in the
solutions of the hyperbolic heat equation (see Nagy, Ortiz, and Reula
[8]), and the studies of the relationship between the relativistic
Navier-Stokes and the causal fluid theories in the near equilibrium
fluid states (see Hiscock and Lindblom [9]).

Let us begin by recalling the theory of a perfect fluid: the
mathematical description of a fluid having negligible internal
dissipation.  The state of such a fluid is determined by three fields on
spacetime: a future directed unit timelike vector field, $u^a$, and two
scalar fields $n$ and $\rho$.  These fields are assumed to be solutions
of the differential equations

$$
\nabla_mN^m = 0,\eqno(1.1)
$$
$$
\nabla_m T^{ma} = 0,\eqno(1.2)
$$
\noindent where $N^a$ and $T^{ab}$ are given in terms of the fluid fields
by
$$
N^a = nu^a,\eqno(1.3)
$$
$$
T^{ab} = (\rho+p)\, u^a u^b + p\, g^{ab}.\eqno(1.4)
$$

\noindent Here $p$ is a smooth function of $n$ and $\rho$ (the equation
of state), that is fixed once and for all for a given type of fluid.
The conserved vector $N^a$ is the particle current of the fluid, and
thus $u^a$ may be identified as the four-velocity and $n$ as the number
density as measured by an observer co-moving with the fluid.  The
conserved tensor $T^{ab}$ is the stress energy of fluid.  Thus from
eq.~\(1.4), $\rho$ is identified as the mass-energy density and $p$ as the
pressure of the fluid, both as measured by a co-moving observer.  These
quantities are all directly observable because the particle current
$N^a$ and the stress energy $T^{ab}$ are themselves directly observable.

  The theory of a perfect fluid, eqs.~\(1.1)--\(1.4), has a number of
attractive mathematical properties.  One of the most important of these
is that eqs.~\(1.1)--\(1.2) form a symmetric-hyperbolic and causal system
when suitable restrictions are placed on the equation of state.  Let
$\xi^\alpha=(n,\rho,u^a)$ denote the dynamical fluid fields.  Then
eqs.~\(1.1)--\(1.2) are equivalent to

$$
M^m{}_{\alpha\beta}\nabla_m\xi^\beta=0,\eqno(1.5)
$$

\noindent where

$$
M^m{}_{\alpha\beta}=P_\alpha{\partial N^m\over \partial \xi^\beta}
+ P_{\alpha a} {\partial T^{am}\over \partial \xi^\beta}.\eqno(1.6)
$$

\noindent The quantities $P_\alpha$ and $P_{\alpha a}$ (functions of
$\xi^\alpha$ and the spacetime metric $g_{ab}$) may be chosen so that
these equations are symmetric in the sense that
$M^m{}_{\alpha\beta}=M^m{}_{\beta\alpha}$ (see Ruggeri and Strumia [10],
Geroch and Lindblom [7], and \S III below).  When suitable
restrictions are placed on the equation of state these equations are
also hyperbolic and causal, because
$\lambda^m=M^m{}_{\alpha\beta}Z^\alpha Z^\beta$ is past directed
timelike for every choice of $Z^\alpha\neq0$ in these theories.

Next, let us turn to the main subject of this paper: theories of
dissipative fluids.  Since there is as yet no universally accepted
theory for such fluids, a rather broad class of theories has been
included in this discussion.  The state of the fluid in these theories
is determined by two sets of fields, $\xi^\alpha$ and $\varphi^A$, each
representing some collection of tensor fields (possibly subject to
certain algebraic constraints) on spacetime.  The $\xi^\alpha$ are to
represent, as in the perfect fluid case, the dynamical fluid fields.
The $\varphi^A$ are to represent additional `dissipation' fields that
are needed to complete an acceptable causal fluid theory.  It seems
reasonable to restrict the dimension of the combined $\xi^\alpha$ and
$\varphi^A$ spaces to be equal to the number of independent observable
fields in the theory.\footnote{${}^1$}{This restriction is not required
in the analysis presented here however.  If the number of fluid fields
were taken to be larger than the number of observables then some of the
fluid fields would not be observable.  The results derived here would
still apply, but some of them would change character from experimentally
testable predictions to mathematical identities.} In the case of a
simple dissipative fluid---the case that is of primary interest
here---the particle current $N^a$ and the stress-energy tensor $T^{ab}$
are the independent observable fields.  Hence the most appropriate
choice for the dimension of these combined spaces is fourteen in this
case.  For now, however, neither the structures nor the dimensions of
these spaces will be restricted.

The fields, $\xi^\alpha$ and $\varphi^A$, are assumed to be solutions of
the system of equations

$$
M^m{}_{\alpha\beta}\nabla_m\xi^\beta
+M^m{}_{\alpha A}\nabla_m\varphi^A = 0,\eqno(1.7)
$$

$$
M^m{}_{AB}\nabla\varphi^B+M^m{}_{\alpha A}\nabla_m\xi^\alpha
= -I_{AB}\varphi^B.\eqno(1.8)
$$

\noindent The quantities $M^m{}_{\alpha\beta}$, $M^m{}_{\alpha A}$,
$M^m{}_{AB}$, and $I_{AB}$ are assumed to be smooth functions, fixed
once and for all for a given theory, of the fields $\xi^\alpha$,
$\varphi^A$, and the spacetime metric $g_{ab}$.  Thus
eqs.~\(1.7)--\(1.8) form a first-order system of partial differential
equations for the fluid fields $\xi^\alpha$ and $\varphi^A$.  Three
conditions are now imposed on this system of equations.  These
conditions are very general and should apply to essentially any theory
of fluids (including those describing superfluids, mixtures of different
kinds of fluids, etc.).

\vskip 6pt \noindent {\twelveit
Condition i)}\kern 1em The first condition is on the $M$'s that appear
on the left sides of eqs.~\(1.7)--\(1.8).  Assume that the $M$'s
are symmetric, $M^m{}_{\alpha\beta}=M^m{}_{(\alpha\beta)}$ and
$M^m{}_{AB}=M^m{}_{(AB)}$; and assume that every vector
$\lambda^m$ given by

$$ \lambda^m = M^m{}_{\alpha\beta}Z^\alpha Z^\beta
+2M^m{}_{\alpha A}Z^\alpha Z^A +M^m{}_{AB}Z^A Z^B,\eqno(1.9)
$$

\noindent for some $(Z^\alpha,Z^A)\neq 0$ is past-directed timelike.
This is just the condition needed to insure that the system
\(1.7)--\(1.8) is symmetric, hyperbolic, and causal (see for example
Geroch and Lindblom [11] or M\"uller and Ruggeri [12]).  \vskip 6pt

\noindent {\twelveit Condition ii)}\kern 1em The second condition
involves the tensor $I_{AB}$ that appears on the right side of
eq.~\(1.8).  Assume that $I_{AB}Z^AZ^B>0$ for every $Z^A\neq0$.${}^2$
This condition is adopted to insure, as will be seen more clearly below,
that this fluid theory is strictly dissipative.  It is precisely
analogous to requiring that the viscosity coefficients and the thermal
conductivity not vanish in the Navier-Stokes equation.  \vskip 6pt

\noindent {\twelveit Condition iii)}\kern 1em The third condition
concerns the conservation laws.  Assume that there exist specific
smooth functions $N^a$ and $T^{ab}$ of the fields $\xi^\alpha$,
$\varphi^A$, and $g_{ab}$, such that eq.~\(1.7) implies the conservation
laws, eqs.~\(1.1) and \(1.2).  This condition merely insures that the
theory possesses a conserved stress energy tensor and particle current.
\vskip 6pt

The\footnote{}{${}^2$ The somewhat more general function
$-I_A(\xi^\alpha,\varphi^B, g_{ab})$ could have been adopted for the
right side of eq.~\(1.8) if it satisfied a few additional constraints.
This more general form is equivalent to that given in eq.~\(1.8) if and
only if $I_A$ satisfies the following three conditions: {\twelveit a)}
$I_A=0$ when $\varphi^B=0$, {\twelveit b)} $I_A\varphi^A>0$ when
$\varphi^B\neq0$, and {\twelveit c)} $\partial I_A/\partial \varphi^B$
is not degenerate when evaluated at $\varphi^C=0$.} main result of this
paper is derived in \S II.  It is shown that physical states of the
fluid relax---on a time scale $\tau$ that is characteristic of the
inter-particle interactions---to ones in which the dissipation field is
determined in effect by the dynamical field $\xi^\alpha$ and its
derivative.  In particular, a bound is derived for the quantity
$\Delta\varphi^A$, defined by

$$
\varphi^A = - \left[\left(I^{-1}\right)^{AB}
M^m{}_{\alpha B}\nabla_m\xi^\alpha\right]_{\varphi^C=0}
+\Delta\varphi^A,
\eqno(1.10)
$$

\noindent in the physical fluid states of any fluid theory which
satisfies Conditions {\twelveit i)--iii)}.  This bound on
$\Delta\varphi^A$ is smaller by the factor $b^{5/2}(\tau v/L)^2$ than it
is expected to be.  The constant $v$ is a characteristic sound speed,
$L$ is a macroscopic length scale that characterizes the particular
state of the fluid, and the constant $b$ is a dimensionless bound on
$M^m{}_{AB}$, $M^m{}_{\alpha A}$, $(I^{-1})^{AB}$ and their derivatives
(which will be defined precisely in \S II).  The constant $b$ is
expected to be of order unity for `reasonable' fluid theories.  Thus the
factor $b^{5/2}(\tau v/L)^2$ should be extremely small for real fluids.
For example in water $(\tau v/L)^2\approx 10^{-12}$ for fluid states
with $L\approx 0.1$cm.  Since the dissipation field in a relaxed fluid
state is determined in effect by the dynamical field $\xi^\alpha$ and
its derivative, then so are all other functions of the fluid fields.  In
particular, the particle current $N^a$ and stress energy tensor $T^{ab}$
are given by

$$
 N^a =\left[N^a - {\partial N^a\over\partial \varphi^A}
\left(I^{-1}\right)^{AB}
M^m{}_{\alpha B}\nabla_m\xi^\alpha\right]_{\varphi^C=0}
+\Delta N^a,
\eqno(1.11)
$$

$$
T^{ab} =
\left[T^{ab} - {\partial T^{ab}\over\partial \varphi^A}
\left(I^{-1}\right)^{AB}
M^m{}_{\alpha B}\nabla_m\xi^\alpha\right]_{\varphi^C=0}+
\Delta T^{ab}.
\eqno(1.12)
$$

\noindent It is shown that the quantities $\Delta N^a$ and $\Delta T^{ab}$
are also smaller than their expected values by the factor $b^3(\tau
v/L)^2$.  These results apply to any dissipative fluid theory that
satisfies Conditions {\twelveit i)--iii)} above, and to any physical
fluid state (i.e., as defined more precisely below, a state in which the
spatial and temporal variations of the fluid fields are larger than the
microscopic scales).  This result explains why the independent dynamics
of the dissipation field $\varphi^A$ is never observed: its value is
determined in effect by the dynamical field $\xi^\alpha$ and its
derivative, via eq.~\(1.10), on any time scale over which a macroscopic
measurement of the system can be made.  Although measurements could in
principle be carried out on fluid systems over very short time and
distance scales, it is not required or even expected that such
measurements will be modeled in detail by any macroscopic fluid theory.

The results of \S II show that a dissipative fluid quickly relaxes to a
state in which the particle current and stress-energy tensor are
determined (in effect) by the dynamical fluid field $\xi^\alpha$ and its
derivative $\nabla_m\xi^\alpha$.  Such relationships are quite familiar
to us; for these are precisely the forms that the expressions for these
quantities take in the Navier-Stokes theory.  Recall that in the
relativistic Navier-Stokes theory (as formulated by Eckart [1]) the
particle current and stress-energy tensor are given in terms of the
fields $\xi^\alpha=(n,\rho,u^a)$ by

$$
N^a=nu^a,\eqno(1.13)
$$

$$
T^{ab} = (\rho+p) u^au^b + pg^{ab} + \tau^{ab}+ \tau q^{ab}
+ 2u^{(a}q^{b)},\eqno(1.14)
$$

\noindent where

$$
\tau^{ab}=2 \eta_1 \left[q^{am} q^{bc} - {1\over 3} q^{ab} q^{cm}\right]
\nabla_{(m} u_{c)},\eqno(1.15)
$$

$$
\tau=\eta_2\nabla_mu^m,\eqno(1.16)
$$

$$
q^a = -\kappa \left(q^{am}\nabla_m T + T u^m\nabla_m u^a\right).\eqno(1.17)
$$

\noindent The quantities $\eta_1$, $\eta_2$, and $\kappa$
(positive functions of $n$ and $\rho$) are the viscosities and thermal
conductivity respectively; and, the quantity $T$ (a function of $n$ and
$\rho$) is the thermodynamic temperature which satisfies the first law
of thermodynamics,

$$
d\rho = nTds + {\rho + p\over n} dn.\eqno(1.18)
$$

\noindent In this theory the conservation laws, eqs.~\(1.1)--\(1.2), and
eqs.~\(1.15)--\(1.17) are the differential equations that determine the
field $\xi^\alpha$.  If the dissipation fields are defined as,
$\varphi^A=(\tau,q^a,\tau^{ab})$, then this theory is of the same
general form as those being considered here.  The conservation laws are
precisely in the form of eq.~\(1.7), while eqs.~\(1.15)--\(1.17) have
the form of eq.~\(1.8).  This theory fails to be an acceptable theory
because $M^m{}_{AB}=0$ and thus it fails to satisfy Condition {\twelveit
i)}.  Note that for this relativistic Navier-Stokes theory, the
quantities $\Delta \varphi^A$, $\Delta N^a$, and $\Delta T^{ab}$ as
defined in eqs.~\(1.10)--\(1.12) vanish identically.

The vanishing (effectively) of $\Delta N^a$ and $\Delta T^{ab}$ for
the general dissipative fluid theories considered here implies that
the particle current and stress-energy tensor depend (in effect) only on
$\xi^\alpha$ and its derivative $\nabla_m\xi^\alpha$ in any physical
fluid state.  In the relativistic Navier-Stokes theory, however, only
certain components of $\nabla_m\xi^\alpha$ appear in these
expressions.  For example, in the Navier-Stokes theory $\nabla_mT$
appears in these expressions but not the gradient of any other
thermodynamic scalar.  It is natural to ask then, what class of fluid
theories have the property that their fluid states always relax to
ones in which the particle current and stress-energy tensor are in
effect indistinguishable from those of the relativistic Navier-Stokes
theory?  Or in particular, in which theories do $N^a$ and $T^{ab}$
depend on $\xi^\alpha$ and $\nabla_m\xi^\alpha$ in precisely the same
way as in the Navier-Stokes theory? The following two
additional conditions are necessary and sufficient to guarantee that a
theory will be indistinguishable from Navier-Stokes in this way:

\vskip 6pt \noindent {\twelveit Condition iv)}\kern 1em The fourth
condition concerns the space of dynamical fields $\xi^\alpha$.  Assume
that eq.~\(1.7) is precisely equivalent to the conservation laws,
eqs.~\(1.1)--\(1.2).  This implies that the space of the $\xi^\alpha$
consists of one vector and one scalar field which may, without loss of
generality (as shown in \S III), be taken to be
${\xi}^\alpha=(n,\rho,u^a)$, where $nu^a=N^a$ (with $u^au_a = -1$) and
$\rho=u_au_bT^{ab}$.

\vskip 6pt \noindent {\twelveit Condition v)}\kern 1em The fifth
condition concerns the tensor $M^m{}_{\alpha A}$ that appears in
eqs.~\(1.7)--\(1.8).  Assume that $M^m{}_{\alpha A} \nabla_m\xi^\alpha$
depends on $\nabla_mn$ and $\nabla_m\rho$ only in the combination
$\nabla_mT=(\partial T/\partial n)_\rho\nabla_mn+(\partial
T/\partial\rho)_n \nabla_m\rho$ in the $\varphi^A=0$ states of the
fluid.  This condition is required to insure that heat flow is generated
by the gradient of the thermodynamic temperature $T$ and not the
gradient of some other thermodynamic scalar.  This condition is
equivalent to the requirement that the equilibrium states of the fluid
be `isothermal.'

\vskip 6pt \noindent The theories that satisfy these two additional
conditions are the natural causal generalizations of the Navier-Stokes
theory: the causal theories of a simple dissipative fluid.

In \S III the expressions for the relaxed forms of the particle current
and stress-energy tensor are evaluated for the theories of simple
dissipative fluids, i.e., those satisfying Conditions {\twelveit
i)--v)}.  With the convenient choice of dynamical fields,
$\xi^\alpha=(n,\rho, u^a)$, eqs.~(1.11)--(1.12) reduce to

$$
N^a = nu^a,\eqno(1.19)
$$

$$
\eqalign{
T^{ab} =& (\rho+p) u^au^b + pg^{ab}
+2 \eta_1 \left[q^{am} q^{bc} - {1\over 3} q^{ab} q^{cm}\right]
\nabla_{(m} u_{c)}+\eta_2 \nabla_mu^m q^{ab}\cr
&\qquad\quad
- 2\kappa u^{(a}\left[q^{b)m}\nabla_mT + T u^{|m|}\nabla_mu^{b)}\right]+
\Delta T^{ab},
\cr}\eqno(1.20)
$$

\noindent for suitably chosen functions (of $n$ and $\rho$) $p$,
$\eta_1$, $\eta_2$, and $\kappa$.  Since $\Delta T^{ab}$ is extremely
small in the physical fluid states of these theories, this shows that
the particle current and stress energy tensor are (in effect)
indistinguishable from those of the relativistic Navier-Stokes
theory.\footnote{${}^3$}{Note that $\Delta N^m$ and $u_au_b\Delta
T^{ab}$ vanish identically as a consequence of the particular choice of
$\xi^\alpha$ made here.  Had a different choice been made, such as the
one traditionally used in the Landau-Lifschitz theory [2], then other
components of these quantities would have vanished identically instead.}

\vskip 1cm \noindent \S II The Relaxation Effect \vskip .5cm

The key result in this paper is that the physical states of the fluid
relax to ones in which the dissipation field $\varphi^A$ is determined
(in effect) by the dynamical fluid field $\xi^\alpha$ and its derivative
$\nabla_m\xi^\alpha$.  That some form of relaxation should occur in the
solutions of eqs.~\(1.7)--\(1.8) can be seen fairly easily.  Consider
the quantity $I_{AB}\varphi^B+M^m{}_{\alpha A}\nabla_m\xi^\alpha$.  If
this quantity does not vanish at some point, then the first term in
eq.~\(1.8) causes $\varphi^A$ to evolve in the direction that tends to
make it vanish.  The rate at which this evolution occurs is determined
by the time scale that is encoded in the tensor $I_{AB}$.  For fluids
this time scale will be determined by the viscosity and thermal
conductivity coefficients contained in $I_{AB}$, and therefore will be
characteristic of the inter-particle interaction times for the fluid.
The demonstration that the quantity $\Delta\varphi^A$ defined in
eq.~\(1.10) is small will be done in two steps.  First, it is shown that
a related quantity $\varphi^A+\Lambda^A$, defined below, is small using
a fairly simple and straightforward argument.  Second, a slightly more
elaborate argument shows that the quantities $\Delta
\varphi^A$, $\Delta N^a$, and $\Delta T^{ab}$ of eqs.~\(1.10)--\(1.12)
are also small.

 Begin by obtaining the following equation for $\varphi^A+\Lambda^A$
from eq.~\(1.8):

$$
\eqalign{
&\nabla_m\left[M^m{}_{AB}\left(\varphi^A+\Lambda^A\right)
\left(\varphi^B+\Lambda^B\right)\right]\cr
&\qquad\qquad\qquad\qquad=-2I_{AB}\left(\varphi^A+\Lambda^A\right)
\left(\varphi^B+\Lambda^B\right)+ \left(\varphi^A+\Lambda^A\right)
{\cal{A}}_A,\cr}\eqno(2.1)
$$

\noindent where $\Lambda^A$ and ${\cal A}_A$
are defined by

$$ \Lambda^A = \left(I^{-1}\right)^{AB}
\left(M^m{}_{\alpha B}\nabla_m\xi^\alpha -{1\over
2}\varphi^C\nabla_mM^m{}_{BC}\right),\eqno(2.2)
$$

$$ {\cal{A}}_A =
\Lambda^B \nabla_m M^m{}_{AB} + 2 M^m{}_{AB} \nabla_m\Lambda^B.
\eqno(2.3)
$$

\noindent Next consider $S(0)$, a bounded open subset of
some Cauchy surface.  Use the timelike vector field whose divergence
appears on the left side of eq.~\(2.1) to define a map between the
points on successive Cauchy surfaces.  Let $S(\kappa)$ denote the image
of $S(0)$ under this map into the Cauchy surface labeled by the time
function $\kappa$.  Choose this time function $\kappa$ so that it
satisfies

$$ I_{AB}Z^AZ^B \geq - Z^A Z^B M^m{}_{AB}\nabla_m
\kappa\eqno(2.4)
$$
\noindent (for every $Z^A$), in the spacetime
region $(0,\kappa_o)\times S(\kappa_o)$.  Next, define the following ${\cal
L}^2$ norm of $\varphi^A+\Lambda^A$,

$$ \alpha^2(\kappa)
=\int_{S(\kappa)} {\cal{G}}_{AB} \left(\varphi^A+\Lambda^A\right)
\left(\varphi^B+\Lambda^B\right)dV,\eqno(2.5)
$$

\noindent where ${\cal G}_{AB} = n_m M^m{}_{AB}$, and $n_m$ is the
future directed unit vector proportional to $\nabla_m \kappa$.  The
evolution of this norm is determined by integrating eq.~\(2.1) over the
spacetime region consisting of points in $S(\kappa)$ that lie between
two nearby $\kappa$ = constant slices.  The integral along the timelike
boundary of this region vanishes because of the choice of $S(\kappa)$.
The integral of the terms on the right in eq.~\(2.1) may be transformed
using eq.~\(2.4) for the first term and the Schwartz inequality for the
second.  Taking the limit as the difference between $\kappa$ on these
two slices goes to zero, the following differential inequality is
obtained for $\alpha$,

$$ {d \alpha\over d\kappa}\leq -\alpha
+{1\over 2}||{\cal{A}}||, \eqno(2.6)
$$

\noindent where

$$
||{\cal{A}}||(\kappa)= \left[\int_{S(\kappa)}{{\cal{G}}^{AB}{\cal{A}}_A
{\cal{A}}_BdV \over -\nabla_m\kappa\nabla^m\kappa}\right]^{1/2},
\eqno(2.7)
$$

\noindent and ${\cal G}^{AB}$ denotes the inverse of
${\cal G}_{AB}$.  This ordinary differential inequality, eq.~\(2.6), can
be integrated to obtain the following bound on $\alpha$,

$$
\alpha(\kappa_o)\leq \alpha(0)e^{-\kappa_o}+ {1\over 2}\int_0^{\kappa_o}
e^{-(\kappa_o-\kappa)} ||{\cal{A}}||(\kappa)d\kappa.  \eqno(2.8)
$$

To proceed further a bound must be obtained for the quantity $||{\cal A}||$
that appears in eq.~\(2.8).  To this end a norm is introduced on tensors:
The positive definite ${\cal G}_{\alpha\beta}
=n_mM^m{}_{\alpha\beta}$ and its inverse ${\cal G}^{\alpha\beta}$ are used for
indices associated with the dynamical field, $\xi^\alpha$; and
the positive definite ${\cal G}_{AB}=n_mM^m{}_{AB}$ and its inverse
${\cal G}^{AB}$ are used for indices associated with the dissipation field
$\varphi^A$.  For spacetime indices the positive definite metric
${\cal G}_{ab}=n_an_b + v^{-2}(g_{ab}+n_an_b)$ and its inverse
$\hat{\cal G}^{ab}=n^an^b + v^{2}(g^{ab}+n^an^b)$ are used.  The constant $v$,
with $0<v<1$, is chosen to be an upper bond on the speed (relative to
$n_a$) of signal propagation, i.e., a number such that $(n_a\lambda^a)^2
\geq v^{-2}(g_{ab}+n_an_b)\lambda^a\lambda^b$ for every $\lambda^a$
given in eq.~\(1.9).  As examples of this norm, the integrand in
eq.~\(2.5) can be written as $|\varphi^A + \Lambda^A|^2 = {\cal
G}_{AB}(\varphi^A+\Lambda^A)(\varphi^B+\Lambda^B)$, while $|{\cal
A}_A|^2={\cal G}^{AB}{\cal A}_A{\cal A}_B$ and
$|\nabla_m\kappa|^2=\hat{\cal G}^{ab}\nabla_a\kappa\nabla_b\kappa
=-\nabla_a\kappa\nabla^a\kappa$.  Note that $|M^m{}_{AB}|\leq 4d$
where $d$ is the dimension of the space of dissipation fields.  Since
$d$ will be some relatively small integer, say $d=9$,
the norm of the $M$'s will be of order unity in these fluid theories.

In the fluid theories considered here the quantities $M^m{}_{AB}$,
$M^m{}_{\alpha A}$ and $I_{AB}$ are assumed to be smooth functions of
the fluid fields.  Therefore, these quantities and their derivatives
with respect to the fluid fields are bounded.  It is convenient to
quantify these bounds in terms of three constants $b$, $\tau$ and
$\zeta$.  Consider fluid fields $\xi^\alpha$ and $\varphi^A$ that are
bounded by the constant $\zeta$:

$$
|\varphi^A|\leq \zeta,\quad |\xi^\alpha|\leq \zeta.  \eqno(2.9)
$$

\noindent Next define the dimensionless constant $b$ to be a bound
on the $M$'s and their derivatives.  In particular assume that

$$
\eqalign{
&\quad\,\,\,|M^m{}_{AB}|\leq b,
\quad \left|{\partial M^m{}_{AB}\over\partial\xi^\alpha}\right|
\leq {b\over\zeta},
\quad \left|{\partial M^m{}_{AB}\over\partial\varphi^C}\right|
\leq {b\over\zeta},\cr
&\left|{\partial^2 M^m{}_{AB}
  \over\partial\xi^\alpha\partial \xi^\beta}\right|
  \leq {b\over\zeta^2},
\quad \left|{\partial^2 M^m{}_{AB}
  \over\partial\xi^\alpha\partial\varphi^C}\right|
  \leq {b\over\zeta^2},
\quad \left|{\partial^2 M^m{}_{AB}
  \over\partial\varphi^C\partial\varphi^D}\right|
  \leq {b\over\zeta^2},\cr}\eqno(2.10)
$$

\noindent and

$$
|M^m{}_{\alpha A}|\leq b,
\quad \left|{\partial M^m{}_{\alpha A}\over\partial\xi^\beta}\right|
\leq {b\over\zeta},
\quad \left|{\partial M^m{}_{\alpha A}\over\partial\varphi^B}\right|
\leq {b\over\zeta}.\eqno(2.11)
$$

\noindent Finally, the constant $\tau$ is defined as a bound on
$(I^{-1})^{AB}$ and its derivatives

$$
|(I^{-1})^{AB}|\leq b\tau,
\quad
\left|{\partial (I^{-1})^{AB}\over \partial \xi^\alpha}\right|\leq
{b\tau\over\zeta},
\quad
\left|{\partial (I^{-1})^{AB}\over \partial \varphi^C}\right|\leq
{b\tau\over\zeta},\eqno(2.12)
$$

\noindent for $b$ given above.
The constant $\tau$ that appears in these bounds is
the characteristic time scale on which the dissipative term $I_{AB}$
influences the evolution of the fluid in eq.~\(1.8).  This constant
also fixes the relationship between physical time and the time
function $\kappa$ because of eq.~\(2.4).  The time function $\kappa$
can be chosen so that

$$
|\nabla_m\kappa| \geq {1\over\tau}.\eqno(2.13)
$$

\noindent This $\kappa$ in effect measures time in units of $\tau$.

To proceed further bounds must now be placed on the spatial derivatives
of the fluid fields.  Assume that there exists a constant $L$ such that

$$
|\nabla_m\xi^\alpha|\leq {v \zeta\over L},
\quad |\nabla_m\varphi^A|\leq {v \zeta\over L},
\quad|\nabla_m\nabla_n\xi^\alpha|\leq {v^2 \zeta\over L^2},
\quad |\nabla_m\nabla_n\varphi^A|\leq {v^2 \zeta\over L^2}.\eqno(2.15)
$$

\noindent These inequalities restrict the solutions to the fluid
equations\footnote{${}^4$}{It is expected that large numbers of
solutions to the fluid equations exist which satisfy these conditions.
In particular, it is expected that initial data satisfying these
conditions on a Cauchy surface will evolve (for some macroscopic time)
as a solution that satisfies these conditions everywhere in the
development of these data.  There do not exist theorems at present,
however, which prove the existence of solutions having these
properties.} to those which do not vary appreciably on length scales
shorter than $L$ and on time scales shorter than $L/v$.  These
inequalities select, therefore, the set of solutions that represent real
physical fluid states.  Fluid states in which rapid variations of the
fluid fields occur on time and length scales smaller than the
microscopic particle interaction scales probably can not be
adequately modeled by any macroscopic fluid theory.  Thus, solutions to
the fluid equations having these properties are not considered physical.
The inequalities in eq.~\(2.15) therefore select out the physical
solutions of the fluid equations when $L$ is larger than the microscopic
interaction length scale.  For these solutions the quantity $||{\cal
A}||$ can be bounded by using eqs.~\(2.10)--\(2.15) in eq.~\(2.7):

$$
||{\cal A}||(\kappa)\leq 26 b^3{\zeta}\left({\tau v\over L}\right)^2
V^{1/2}(\kappa),\eqno(2.16)
$$

\noindent where $V(\kappa)$ is the volume of the region $S(\kappa)$:

$$
V(\kappa)=\int_{S(\kappa)}dV.\eqno(2.17)
$$

\noindent Including this bound into eq.~\(2.8) the following
bound is then obtained for $\alpha$:

$$
\alpha(\kappa_o)\leq\alpha(0)e^{-\kappa_o} + 13 b^3 \zeta
\left({\tau v\over L}\right)^2 \int_0^{\kappa_o}
e^{-(\kappa_o-\kappa)}V^{1/2}(\kappa)d\kappa.
\eqno(2.18)
$$

\noindent This bound consists of two pieces.  The first is simply the
initial value of $\alpha$ multiplied by $e^{-\kappa_o}$.  This term
falls exponentially to zero on the characteristic time scale $\tau$.
The second term is a constant multiplied by the time average of the
spatial volume over which the norm $\alpha$ is defined.  This second
term, the asymptotic bound on the norm $\alpha$, is smaller than the
{\twelveit a priori} expectation of its value, $\zeta (bV)^{1/2}$, by
the factor $b^{5/2}(\tau v/L)^2$.  This factor will be extremely
small, being proportional to the square of the ratio of the
characteristic dissipation time scale $\tau$ to the characteristic
dynamical time scale $L/v$, as long as the constant $b$ is of order
unity and the constant $v$ is comparable to the sound speed in the
material.  The constant $b$ is a measure of the $M$'s and $I$ and
their derivatives with respect to the fluid fields.  This constant
will be of order unity unless these quantities depend on the fluid
fields in a very perverse way (e.g. if the $M$'s dependence on the
fields were highly oscillatory).  The constant $v$ will be comparable
to the sound speed of the material as long the foliation of Cauchy
surfaces is chosen so that the fluid motion is not highly supersonic,
and as long as the characteristic speeds associated with the
dissipation fields are comparable to the usual sound speed.  In this
case the bound on $\alpha$ derived in eq.~\(2.18) implies that the
dissipation field $\varphi^A$ relaxes in the physical states of these
fluid theories in such a way that the quantity $\varphi^A +\Lambda^A$
becomes extremely small.

To complete the argument that the quantities $\Delta \varphi^A$, $\Delta
N^a$, and $\Delta T^{ab}$ of eqs.~\(1.10)--\(1.12) are small, additional
${\cal L}^2$ and ${\cal L}^4$ bounds are needed on the dissipation field
$\varphi^A$.  To obtain these bounds the following
identities are derived from eq.~\(1.8),

$$
\nabla_m\left[M^m{}_{AB}\varphi^A\varphi^B\right]=
-2I_{AB}\varphi^A\varphi^B+
\varphi^A{\cal{B}}_A,\eqno(2.19)
$$

$$
\nabla_m\left[M^m{}_{AB}{\cal{G}}_{CD}\varphi^A\varphi^B
\varphi^C\varphi^D\right]=
-2I_{AB}{\cal G}_{CD}\varphi^A\varphi^B\varphi^C\varphi^D
+\varphi^A\varphi^B\varphi^C{\cal C}_{ABC},\eqno(2.20)
$$

\noindent where ${\cal B}_A$ and ${\cal C}_{ABC}$ are defined by

$$
{\cal{B}}_A = \varphi^B\nabla_mM^m{}_{AB}
-2M^m{}_{\alpha A}\nabla_m\xi^\alpha,\eqno(2.21)
$$

$$
{\cal{C}}_{ABC} = {\cal G}_{AB}{\cal B}_C
+2 M^m{}_{AB}{\cal G}_{CD}\nabla_m\varphi^D
+M^m{}_{AB}\varphi^D\nabla_m{\cal G}_{CD}.
\eqno(2.22)
$$

\noindent Next an integral norm, analogous to $\alpha$ above, is defined
for the field $\varphi^A$:

$$
\beta^2(\kappa)= \int_{\hat{S}(\kappa)}\left[
K_1^2 \left|\varphi^A+\Lambda^A\right|^2
+K_2^2\left|\varphi^A\right|^2
+K_3^2\left|\varphi^A\right|^4
\right]dV,\eqno(2.23)
$$

\noindent where $K_1>0$, $K_2>0$, and $K_3>0$ are constants whose values
will be specified later.  The time evolution of $\beta$ is determined in
analogy with eq.~\(2.6) by integrating $K_1^2$ multiplied by eq.~\(2.1),
plus $K_2^2$ multiplied by eq.~\(2.19), plus $K_3^2$ multiplied by
eq.~\(2.20), over the spacetime region consisting of points
in $\hat{S}(\kappa)$ that lie between two nearby $\kappa$ =
constant slices.  The sequence of spatial sections $\hat{S}(\kappa)$ is
chosen in this case so that the timelike boundary integral vanishes
identically here as well.  The integrations on the right side of this
equation may be simplified, again in analogy with eq.~\(2.6), by using
eq.~\(2.4) and the Schwartz inequality.  The result is the following
differential inequality on $\beta$,

$$
{d\beta\over d\kappa} \leq - \beta + {1\over 2}\left( K_1 ||{\cal{A}}||
+ K_2 ||{\cal{B}}||\right)
+{1\over 2}K_3\left[\int_{\hat{S}(\kappa)}
{\left|{\cal C}_{ABC}\right|^2\left|\varphi^D\right|^2 dV \over
\left|\nabla_m\kappa\right|^2}\right]^{1/2},
\eqno(2.24)
$$

\noindent where $||{\cal A}||$ is given by eq.~\(2.7) and $||{\cal B}||$
is

$$
||{\cal{B}}||(\kappa)=
\left[\int_{\hat{S}(\kappa)}{\left|{\cal{B}}_A\right|^2dV \over
\left|\nabla_m\kappa\right|^2}\right]^{1/2}.
\eqno(2.25)
$$

\noindent The quantities that appear on the right side of eq.~\(2.24) can
be bounded if one additional restriction is made on the physical
solutions of the fluid theory.  Assume that the extrinsic curvature
and acceleration of the $\kappa\,\,\,=$ constant surfaces are bounded by

$$
|\nabla_an_b|\leq {v\over L}.\eqno(2.26)
$$

\noindent Using the bounds given in eqs.~\(2.9)--\(2.13), \(2.15),
and \(2.26), the following bounds can be obtained for the quantities
that appear on the right side of eq.~\(2.24):

$$
||{\cal B}||\leq 4b\zeta {\tau v\over L}\hat{V}^{1/2}(\kappa),
\eqno(2.27)
$$

$$
\left[\int_{\hat{S}(\kappa)}
{\left|{\cal C}_{ABC}\right|^2\left|\varphi^D\right|^2
dV\over |\nabla_m\kappa|^2}\right]^{1/2}
\leq 9b^2\zeta{\tau v\over L}{\beta\over K_2}.
\eqno(2.28)
$$

\noindent Combining these bounds with eq.~\(2.16), the differential
inequality for $\beta$ can be simplified to the following

$$
{d\beta\over d\kappa}\leq -\left[1-{9\over 2}b^2\zeta{\tau v\over L}
{K_3\over K_2}\right]\beta
+\left[13b^2\left({\tau v\over L}\right)K_1
+ 2K_2\right]b\zeta \left({\tau v\over L}\right)\hat{V}^{1/2}(\kappa).
\eqno(2.29)
$$

\noindent  The constants $K_1$, $K_2$ and $K_3$ are now chosen to be

$$
K_1={1\over 13b^2},\quad K_2={1\over 2}\left({\tau v\over L}\right),
\quad K_3={1\over 18b^2\zeta}.
\eqno(2.30)
$$

\noindent With these choices eq.~\(2.29) becomes

$$
{d\beta\over d\kappa}\leq -{\beta\over 2} + 2b
\zeta\left({\tau v\over L}\right)^2\hat{V}^{1/2}(\kappa).
\eqno(2.31)
$$

\noindent Integrating this inequality, the desired bound
on the norm $\beta$ is obtained:

$$
\beta(\kappa_o)\leq \beta(0)e^{-\kappa_o/2}+
4b\zeta\left({\tau v\over L}\right)^2
<\hat{V}^{1/2}>,\eqno(2.32)
$$

\noindent where $<\hat{V}^{1/2}>$ denotes the time average of
the spatial volume,

$$
<\hat{V}^{1/2}>={1\over 2}\int_0^{\kappa_o}e^{-(\kappa_o-\kappa)/2}
\hat{V}^{1/2}(\kappa)d\kappa.
\eqno(2.33)
$$

\noindent This bound on $\beta$ implies an ${\cal L}^2$ bound on
$\varphi^A+\Lambda^A$, and simultaneously ${\cal L}^2$ and ${\cal L}^4$
bounds on $\varphi^A$.  The asymptotic values of these bounds are given
by

$$
\left[\int_{\hat{S}(\kappa)}|\varphi^A+\Lambda^A|^2dV\right]^{1/2}\leq
52b^3\zeta\left({\tau v\over L}\right)^2<\hat{V}^{1/2}>,
\eqno(2.34)
$$

$$
\left[\int_{\hat{S}(\kappa)}|\varphi^A|^2dV\right]^{1/2}\leq
8b\zeta\left({\tau v\over L}\right)<\hat{V}^{1/2}>,
\eqno(2.35)
$$

$$
\left[\int_{\hat{S}(\kappa)}|\varphi^A|^4dV\right]^{1/2}\leq
72b^3\zeta^2\left({\tau v\over L}\right)^2
<\hat{V}^{1/2}>.
\eqno(2.36)
$$

\noindent These bounds on $\varphi^A$ are smaller by the factor $\tau
v/L$ than their {\twelveit a priori} expected values.  The bound on
$\varphi^A+\Lambda^A$ is even smaller, however, being reduced from its
{\twelveit a priori} expected value by the factor $(\tau v/L)^2$.  Thus
$\varphi^A+\Lambda^A$ becomes small not simply because $\varphi^A$ and
$\Lambda^A$ get small individually.  Rather, this quantity becomes small
because $\varphi^A$ approaches $\Lambda^A$ asymptotically.  Note that
the region $\hat{S}(\kappa)$ over which these norms are computed may be
chosen arbitrarily on any particular slice.\footnote{${}^5$}{\noindent
The regions $\hat{S}(\kappa)$ on the other slices in the foliation are
then fixed, however, in order to eliminate the spatial
boundary terms from the integration.}  Also note that
$<\hat{V}^{1/2}(\kappa)>\approx\hat{V}^{1/2}(\kappa)$ if
$\hat{V}(\kappa)\gg(\tau v)^3$.  The time average used here is
exponentially weighted and hence only those slices within about one
microscopic interaction time $\tau$ of $\kappa$ contribute
significantly.

The main results of this section are bounds on the quantities
$\Delta\varphi^A$, $\Delta N^a$, and $\Delta T^{ab}$ defined in
eqs.~\(1.10)--\(1.12).  These bounds are obtained beginning with
the quantity $\Delta\varphi^A$,

$$
\Delta\varphi^A=\varphi^A+\left[(I^{-1})^{AB}M^m{}_{\alpha B}
\nabla_m\xi^\alpha\right]_{\varphi^C=0}.\eqno(2.37)
$$

\noindent This quantity can be re-written as the sum of
$\varphi^A+\Lambda^A$, a quantity whose bound was established above,
plus $\epsilon^A$:

$$
\Delta\varphi^A=\varphi^A+\Lambda^A+\epsilon^A
,\eqno(2.38)
$$

\noindent where $\epsilon^A$ may be written (using the standard
expression for the remainder in a Taylor expansion) as

$$\eqalign{
\epsilon^A=&{1\over 2}(I^{-1})^{AB}\varphi^C\nabla_mM^m{}_{BC}\cr
&\quad-\varphi^C\nabla_m\xi^\alpha
\int_0^1\left\{ {\partial\over\partial\varphi^C}
\left[(I^{-1})^{AB}M^m{}_{\alpha B}\right](\xi^\beta,\lambda\varphi^D)
\right\}d\lambda.\cr}
\eqno(2.39)
$$

\noindent Now, using eqs.~\(2.10)--\(2.12) and \(2.15) it is straightforward
to obtain the following bound on $\epsilon^A$,

$$
|\epsilon^A|\leq 3b^2 \left({\tau v\over L}\right)|\varphi^A|.
\eqno(2.40)
$$

\noindent  Using the triangle inequality for ${\cal L}^2$ norms,
the norm of $\Delta \varphi^A$ can be expressed as the sum of the
norms for $\varphi^A+\Lambda^A$, from \(2.34), and the norm of $\epsilon^A$,
using \(2.35) and \(2.40):

$$
\left[\int_{\hat{S}(\kappa)}|\Delta\varphi^A|^2dV\right]^{1/2}
\leq 38b^3\zeta \left({\tau v\over L}\right)^2
<\hat{V}^{1/2}>.\eqno(2.41)
$$

\noindent  Thus, the norm of $\Delta\varphi^A$ is smaller than its
{\twelveit a priori} expected value by the factor $b^{5/2}(\tau v/L)^2$.

Turn next to the quantity $\Delta T^{ab}$,

$$
\Delta T^{ab}=T^{ab}-\left[T^{ab}- {\partial
T^{ab}\over\partial\varphi^A} (I^{-1})^{AB}M^m{}_{\alpha
B}\nabla_m\xi^\alpha\right]_{\varphi^C=0}. \eqno(2.42)
$$

\noindent This quantity may be re-written (using eq.~\[2.37] and again the
standard expression for the remainder in a Taylor expansion) as

$$
\Delta T^{ab}=\Delta\varphi^A
\left[{\partial T^{ab}\over \partial \varphi^A}\right]_{\varphi^C=0}
+ \varphi^A\varphi^B
\int_0^1\left\{ (1-\lambda){\partial^2 T^{ab}\over\partial
\varphi^A\partial\varphi^B}
(\xi^\alpha,\lambda\varphi^C)\right\}d\lambda.
\eqno(2.43)
$$

\noindent  The norm of this quantity can easily be bounded by

$$
|\Delta T^{ab}|\leq {\epsilon\over\zeta}|\Delta\varphi^A|
+ {\epsilon\over 2\zeta^2}|\varphi^A|^2,
\eqno(2.44)
$$

\noindent if the field derivatives of $T^{ab}$
satisfy the following bounds

$$
\left|{\partial T^{ab}\over\partial\varphi^A}\right|\leq {\epsilon\over\zeta},
\quad
\left|{\partial^2 T^{ab}\over\partial\varphi^A\partial\varphi^B}\right|\leq
{\epsilon\over\zeta^2}.\eqno(2.45)
$$

\noindent The constant $\epsilon$ is a characteristic internal
energy density.  Using the expressions for the bound on
$\Delta\varphi^A$ from eq.~\(2.41) and the bound on $|\varphi^A|^2$
from eq.~\(2.36), the following bound on $\Delta T^{ab}$ is obtained,

$$
\left[\int_{\hat{S}(\kappa)}|\Delta T^{ab}|^{2}dV\right]^{1/2}\leq
112b^3
\left({\tau v\over L}\right)^2\epsilon<\hat{V}^{1/2}>.\eqno(2.46)
$$

\noindent  This equation  provides a bound on
$\Delta T^{ab}$ that is smaller than its {\twelveit a priori} expected
value, $\epsilon <\hat{V}^{1/2}>$, by the factor $b^3(\tau v/ L)^2$.

An exactly analogous bound can be obtained for $\Delta N^a$ if
the field derivatives of $N^a$ are bounded by

$$
\left|{\partial N^{a}\over\partial\varphi^A}\right|\leq {\nu\over\zeta},
\quad
\left|{\partial^2 N^{a}\over\partial\varphi^A\partial\varphi^B}\right|\leq
{\nu\over\zeta^2},\eqno(2.47)
$$

\noindent where $\nu$ is a characteristic number density.  The bound on
$\Delta N^a$ is obtained in precisely the same way as the
bound on $\Delta T^{ab}$, with the result

$$
\left[\int_{\hat{S}(\kappa)}|\Delta N^{a}|^{2}dV\right]^{1/2}\leq
112b^3
\left({\tau v\over L}\right)^2\nu<\hat{V}^{1/2}>.\eqno(2.48)
$$

\noindent  Thus the bound on $\Delta N^a$ is also smaller than
its {\twelveit a priori} expected value by the factor $b^3(\tau v/L)^2$.
\vskip 1cm
\noindent \S III Simple Dissipative Fluids
\vskip .5cm

In this section the relaxed expressions for the particle current and
stress energy tensor, eqs.~\(1.11)--\(1.12), are evaluated for the
theories of a simple dissipative fluid.  Condition~{\twelveit iv)}\kern
.5em guarantees that eq.~\(1.7) is equivalent to the conservation laws
in this case.  This implies that the space of the $\xi^\alpha$ must
consist of one vector and one scalar field.  The form of
eqs.~\(1.7)--\(1.8) is unchanged if the fluid fields are transformed in
the following way: $\hat{\xi}^\alpha =\hat{\xi}^\alpha({\xi}^\beta,
{\varphi}^B)$ and $\hat{\varphi}^A = \hat{\varphi}^A({\varphi}^B)$.  The
choice $\hat{\xi}^\alpha=(n,\rho,u^a)$ and $\hat{\varphi}^A= \varphi^A$,
where

$$
nu^a = N^a(\xi^\beta,\varphi^B,g_{bc}),\eqno(3.1)
$$

$$
\rho = u_au_bT^{ab}(\xi^\beta,\varphi^B,g_{cd}),\eqno(3.2)
$$

\noindent is a transformation of this form.  Thus $\xi^\alpha$ may be
chosen to be $\xi^\alpha=(n,\rho,u^a)$, without loss of generality.
With this choice $\partial N^a/\partial \varphi^A=0$ and $u_au_b\partial
T^{ab}/ \partial \varphi^A =0$.  Evaluating eq.~\(1.11) for this case we
obtain $N^a = nu^a+\Delta N^a$, hence eq.~\(1.19).  The quantity $\Delta
N^a$ vanishes identically as a consequence of the choice of $\xi^\alpha$
used here.  Condition {\twelveit iv)}\kern .5em also implies that the
tensors $M^m{}_{\alpha\beta}$ and $M^m{}_{\alpha A}$ of eq.~\(1.7) must
be given by

$$
M^m{}_{\alpha\beta}=P_\alpha{\partial N^m\over \partial \xi^\beta}
+ P_{\alpha a} {\partial T^{am}\over \partial \xi^\beta},\eqno(3.3)
$$

$$
M^m{}_{\alpha A}=
P_{\alpha a} {\partial T^{am}\over \partial \varphi^A},\eqno(3.4)
$$

\noindent where $P_\alpha$ and $P_{\alpha a}$ are suitably chosen
functions of $\xi^\alpha$, $\varphi^A$ and $g_{ab}$.  Note that the
term proportional to $P_\alpha$ is missing from eq.~\(3.4) because
$\partial N^a/\partial\varphi^ A=0$ for our choice of $\xi^\alpha$.
Using this expression for $M^m{}_{\alpha a}$, the general expression
for $T^{ab}$ in eq.~\(1.12) reduces to

$$
T^{ab} =
\left[T^{ab} - {\partial T^{ab}\over\partial \varphi^A}
\left(I^{-1}\right)^{AB}
{\partial T^{cm}\over\partial\varphi^B}
P_{\alpha c}\nabla_m\xi^\alpha\right]_{\varphi^C=0}+\Delta T^{ab}.
\eqno(3.5)
$$

Condition {\twelveit v)}\kern .5em places restrictions on the allowed
forms of $M^m{}_{\alpha A}\nabla_m\xi^\alpha$ in the fluid states where
$\varphi^A=0$.  From eq.~\(3.4) it follows that this quantity is determined
by $P_{\alpha a}$.  In the $\varphi^A=0$ fluid states the tensor
$M^m{}_{\alpha\beta}$ is identical to the tensor that governs the evolution
of a perfect fluid via eqs.~\(1.5)--\(1.6).  The most general $P$'s that
make $M^m{}_{\alpha\beta}$ symmetric in this case are given by

$$
P_\alpha d\xi^\alpha = -Q_1\left[\left({\partial p\over \partial
n}\right)_\rho^2 +Q_2 \left({\rho+p\over n}\right)^2\right] dn
-Q_1\left[\left({\partial p\over \partial \rho}\right)_n \left({\partial
p\over \partial n}\right)_\rho -Q_2{\rho+p\over n}\right] d\rho,\eqno(3.6)
$$

$$
\eqalign{
P_{\alpha a} d\xi^\alpha
&= u_aQ_1\left[\left({\partial p\over \partial \rho}\right)_n
\left({\partial p\over \partial n}\right)_\rho
-Q_2 {\rho+p\over n}\right] dn
+u_a Q_1\left[\left({\partial p\over \partial \rho}\right)_n^2
+Q_2\right] d\rho\cr
&\,\qquad\qquad\qquad\quad
-Q_1(\rho+p)\left({\partial p\over\partial\rho}\right)_s
q_{ab}
du^b,\cr}\eqno(3.7)
$$

\noindent where $Q_1$ and $Q_2$ are arbitrary functions of $n$ and
$\rho$, and $q_{ab}=g_{ab}+u_au_b$ (see Geroch and Lindblom [7]).  The
hyperbolicity and causality conditions for $M^m{}_{\alpha\beta}$ in this
case are simply, $Q_1>0$ and $Q_2>0$, and the equation of state must
satisfy

$$
0< \left({\partial p\over\partial\rho}\right)_s
\leq 1,
\eqno(3.8)
$$

\noindent with $n>0$ and $\rho+p>0$.  Thus the tensors
$M^m{}_{\alpha\beta}$ and $M^m{}_{\alpha A}$ are determined completely
(up to the arbitrary overall factor $Q_1$) in these fluid states
by the function $Q_2$.

Condition {\twelveit v)} \kern .5em fixes $Q_2$ by demanding that
$M^m{}_{\alpha A}\nabla_m\xi^\alpha$ and hence $P_{\alpha
(a}\nabla_{m)}\xi^\alpha$ depend on $\nabla_mn$ and $\nabla_m\rho$ only
in the combination $\nabla_mT=(\partial T/\partial
n)_\rho\nabla_mn+(\partial T/\partial\rho)_n \nabla_m\rho$.  The unique
$Q_2$ which insures this is

$$
Q_2={1\over nT} \left({\partial p\over \partial \rho}\right)_n
\left({\partial T\over \partial s}\right)_p
\left({\partial p\over \partial T}\right)_s
={1\over n^2T^2}
\left({\partial p\over \partial \rho}\right)_s
\left({\partial \rho\over \partial s}\right)_p
\left({\partial p\over \partial s}\right)_\Theta
,\eqno(3.9)
$$

\noindent where $T$ and $s$ are the temperature and entropy that satisfy
the first law of thermodynamics, eq.~\(1.18), and $\Theta=(\rho+p)/nT-s$.
The second equality in eq.~\(3.9) shows that the condition $Q_2>0$,
needed to insure hyperbolicity of the equations, is equivalent to a well
known condition for thermodynamic stability (see Hiscock and Lindblom
[13]).  With this choice of $Q_2$ the quantity $P_{\alpha
(a}\nabla_{m)}\xi^\alpha$ reduces to

$$
P_{\alpha (a}\nabla_{m)}\xi^\alpha=
Q_1 {\rho + p\over T} \left({\partial p\over \partial \rho}\right)_s
\left[u_{(a}\nabla_{m)}T
- T
\nabla_{(m}u_{a)}\right].\eqno(3.10)
$$

\noindent   The tensor
$\partial T^{ab}/\partial\varphi^A (I^{-1})^{AB} \partial
T^{cd}/\partial\varphi^B$ that appears in eq.~\(3.5) depends only
on $\xi^\alpha$ and $g_{ab}$.
The most general such tensor
(having the appropriate symmetries, etc.) depending only
on $\xi^\alpha$ and $g_{ab}$ is given by

$$
\eqalign{
{\partial T^{ab}\over\partial\varphi^A} \left(I^{-1}\right)^{AB}
{\partial T^{cd}\over\partial\varphi^B}=
{1\over Q_1(\rho+p)} \left({\partial\rho\over\partial p}\right)_s
\Biggl\{&2\eta_1 \left[q^{a(c}q^{d)b}-{1\over 3}q^{ab}q^{cd}\right]\cr
&\quad+\eta_2 q^{ab}q^{cd}-
2\kappa T u^{(a}q^{b)(c}u^{d)}
\Biggr\}.\cr}
\eqno(3.11)
$$

\noindent The arbitrary functions $\eta_1$, $\eta_2$, and $\kappa$ (of
$n$ and $\rho$) that appear in eq.~\(3.11) must be positive as a
consequence of the positivity of $I_{AB}$, from Condition {\twelveit
ii)}, and the positivity of $Q_1$ and $(\partial p/\partial\rho)_s$,
from the hyperbolicity and causality of the equations.\footnote{${}^6$}
{The only requirement on the dissipation fields needed to obtain
eq.~\(3.11) is that the space of the $\varphi^A$ be large enough to
insure that none of the coefficients $\eta_1$, $\eta_2$, or $\kappa$
vanishes identically.  This requires in particular that this space be at
least as large as the nine-dimensional space of symmetric trace-free
tensors.  This is precisely the dimension that is appropriate for a
theory in which the particle current $N^a$ and stress energy tensor
$T^{ab}$ are the only independent observable fields.} Combining this
expression, eq.~\(3.11), with eq.~\(3.10) in eq.~\(3.5) results in the
desired form, eq.~\(1.20).  Thus, the relaxed form of the stress energy
tensor is indistinguishable in these general causal theories from that
of the relativistic Navier-Stokes theory.
\vskip 1cm {\noindent \S IV Concluding Remarks} \vskip .5cm

The argument presented here demonstrates that a relaxation effect takes
place in virtually every causal theory of dissipative fluids.  In the
relaxed fluid states the stress energy tensor and particle current are
well described by expressions that depend only on a subset of the fluid
fields (referred to here as dynamical fluid fields) and their
derivatives.  For those theories that represent simple dissipative
fluids, these expressions are identical to the ones given by the
relativistic Navier-Stokes theory.  This implies that any measurement of
the stress-energy tensor or particle current in these theories (made on
any time and length scale that exceeds the microscopic particle
interaction scales) will give results that are in effect
indistinguishable from those of the Navier-Stokes theory.  Of course the
Navier-Stokes theory is not really a proper physical theory at all since
it is non-causal, unstable, etc.  It is incapable of predicting the
future evolution of initial fluid states.  The argument presented here
shows, nevertheless, that the evolution of {\it any} physical fluid
state according to {\it any} causal theory results in stress-energy
tensors and particle currents that are experimentally indistinguishable
from the Navier-Stokes expressions for these quantities.  Further, this
argument shows that the independent dynamics associated with the
dissipation fields of the fluid (i.e., those additional fluid fields
that are added to the theory to make it causal) is not directly
observable in the physical fluid states.  On a time scale that is
characteristic of the inter-particle interaction times, these
dissipation fields evolve to a relaxed state in which they are
determined in effect by the dynamical fields and their derivatives.

A number of technical improvements could be made to strengthen the
arguments presented here.  The physical fluid states for which this
result applies are those whose gradients are bounded locally to insure
that they are not rapidly changing on microscopic scales.  These local
constraints are much stronger than are actually needed to complete the
proof.  All that is really needed are the ${\cal L}^2$ bounds on the
fluid fields and their derivatives implicit in eqs.~\(2.16), \(2.27),
and \(2.28).  These bounds could undoubtedly be derived using far
weaker ${\cal L}^2$ conditions on the fluid fields and their
derivatives than the local conditions used here.  A more serious
limitation of the present work is its failure to demonstrate the
existence of any solutions at all of the fluid equations which satisfy
these conditions.  The expectation is that essentially every
`physically relevant' solution to the fluid equations does satisfy
these conditions.  In particular it is expected that `almost all'
initial data which are suitably slowly varying on the relevant
microscopic length and time scales will evolve in such a way that
these conditions are preserved for some amount (large on microscopic
scales) of time.  At present, however, theorems of this sort do
not exist for these theories.

Shock waves are one class of physical phenomena that do violate the
conditions imposed on the fluid states in this work.  Significant
differences probably do exist in the descriptions of this type of
fluid phenomenon among the various causal theories and the non-causal
Navier-Stokes equations.  Can meaningful experimental differentiation
among the various theories be found by observing shock waves?  Or, do
the predictions of all macroscopic fluid theories become meaningless
when applied to shocks, since these fluid states all contain rapid
variations on microscopic particle interaction scales?

\vskip .5cm
{\noindent Acknowledgment}
\vskip .5cm

This research was done in collaboration with Robert Geroch;
however, we have written separate accounts of it (see Geroch [14]).
My research was supported by grants from the National Science
Foundation (PHY-9019753) and from the Centre National de la Recherche
Scientifique (UPR 176).
\filbreak
\everypar{\parindent .1cm\hangindent .5cm\hangafter 1}

\vskip .5cm
{\noindent References}
\vskip .5cm

\noindent

\item{1.} C. Eckart, {\sl Phys. Rev.}, {\bf 58}, 919 (1940).

\item{2.} L. Landau, and E. M. Lifschitz, {\sl Fluid Mechanics} (Addison-
Wesley: Reading, MA), Sec. 127 (1975).

\item{3.} W. A. Hiscock, and L. Lindblom, {\sl Phys. Rev.}, {\bf D 31},
725 (1985).

\item{4.} W. Israel, and J. M. Stewart, {\sl Proc. R. Soc. London},
{\bf A 357}, 59 (1979).

\item{5.} B. Carter, in {\sl Relativistic Fluid Dynamics}, ed. by A. Anile,
and Y. Choquet-Bruhat (Springer-Verlag, Berlin), 1 (1989).

\item{6.} I. S. Liu, I. M\"uller, and T. Ruggeri, {\sl Ann. Phys. (N.Y.)},
{\bf 169}, 191 (1986).

\item{7.} R. Geroch, and L. Lindblom, {\sl Phys. Rev.}, {\bf D 41}, 725 (1990).

\item{8.} G. B. Nagy, O. E. Ortiz, O. Reula, {\sl J. Math. Phys.}, {\bf 35},
4334 (1994).

\item{9.} W. A. Hiscock, and L. Lindblom, {\sl Phys. Rev.}, {\bf D 35},
3723 (1987).

\item{10.} T. Ruggeri, and A. Strumia, {\sl Ann. Inst. Henri Poincar\'e A},
{\bf 34}, 65 (1981).

\item{11.} R. Geroch, and L. Lindblom, {\sl Ann. Phys. (N.Y.)}, {\bf
207}, 394 (1991).

\item{12.} I. M\"uller, and T. Ruggeri, {\sl Extended Thermodynamics},
(Springer-Verlag, Berlin) (1993).

\item{13.} W. A. Hiscock, and L. Lindblom, {\sl Ann. Phys. (N.Y.)},
{\bf 151}, 466 (1983).

\item{14.} R. Geroch, {\sl J. Math. Phys.}, in press (1995).

\bye